\documentclass[aps,prd,twocolumn,showpacs,superscriptaddress,preprintnumbers,floatfix,amsmath,amssymb,eufrak,table]{revtex4-2}
\pdfoutput=1
\usepackage{graphicx}
\usepackage{dcolumn}
\usepackage{bm}
\usepackage{slashed}
\usepackage{xcolor}
\usepackage{amsmath,graphicx}
\usepackage[colorlinks=true,linktocpage=true,linkcolor=blue,citecolor=blue]{hyperref}
\usepackage{float}
\usepackage{nicefrac}
\usepackage[normalem]{ulem}
\usepackage{amsmath}
\usepackage{subfigure} 
\usepackage{bbold} 
\usepackage[makeroom]{cancel}
\usepackage{stmaryrd} 

\begin{document}

\title{Charm quark evolution in the early stages of heavy-ion collisions}

\author{Mayank Singh}
\email{mayank.singh@vanderbilt.edu}
\affiliation{Department of Physics and Astronomy, Vanderbilt University, Nashville, Tennessee 37240, USA}

\author{Manu Kurian}
\email{manukurian@iitism.ac.in}
\affiliation{Department of Physics, Indian Institute of Technology (Indian School of Mines) Dhanbad, Jharkhand 826004, India}

\author{Bj\"orn Schenke}
\email{bschenke@bnl.gov}
\affiliation{Physics Department, Brookhaven National Laboratory, Upton, New York 11973, USA}

\author{Sangyong Jeon}
\email{sangyong.jeon@mcgill.ca }
\affiliation{Department of Physics, McGill University, 3600 University Street, Montreal, Qu\'ebec H3A 2T8, Canada}

\author{Charles Gale}
\email{charles.gale@mcgill.ca }
\affiliation{Department of Physics, McGill University, 3600 University Street, Montreal, Qu\'ebec H3A 2T8, Canada}


\begin{abstract}

{Heavy quarks are predominantly generated at the initial stage of relativistic heavy-ion collisions such that heavy flavor observables have the potential to provide information on the pre-equilibrium medium dynamics. In this study, we investigate the sensitivity of D-meson $R_{AA}$ and $v_2$ to early-time charm quark dynamics in Pb+Pb collisions at $\sqrt{s_{NN}}=5.02$ TeV. We employ the IP-Glasma+MUSIC+UrQMD framework to model the evolution of the bulk medium. Charm quarks are generated using PYTHIA with nuclear parton distribution functions and evolved using Langevin dynamics within MARTINI. We observe that even though there is significant momentum broadening in the earliest stage, D-meson $R_{AA}$ and $v_2$ are only weakly sensitive to pre-equilibrium interactions.}

\end{abstract}

\maketitle
\section{Introduction}
Heavy-ion collision experiments conducted at the Relativistic Heavy Ion Collider (RHIC) and the Large Hadron Collider (LHC) offer a unique opportunity to study QCD matter under extreme but controlled conditions. Model-to-data comparisons are the main tool for inferring the properties of the Quantum Chromodynamics (QCD) matter produced. However, modeling heavy-ion collisions is complex and requires a sophisticated multistage framework. The current state-of-the-art approach incorporates initial states with information about the nucleonic and sub-nucleonic degrees of freedom~\cite{Moreland:2014oya,Schenke:2012wb}, hydrodynamical evolution of the Quark-Gluon Plasma (QGP)~\cite{Gale:2013da, DerradideSouza:2015kpt, Romatschke:2017ejr}, and hadronic transport~\cite{Bass:1998ca, SMASH:2016zqf}. The multi-stage model parameters can be constrained using experimental data by performing Bayesian analyses~\cite{Bernhard:2019bmu, JETSCAPE:2020shq, Nijs:2020roc, Heffernan:2023gye}.

Heavy quarks, the charm and bottom quarks (top quarks are rarely produced in the considered collisions), serve as valuable probes for studying the properties of the QCD medium~\cite{vanHees:2005wb,Das:2009vy,Andronic:2015wma,Aarts:2016hap, Das:2024vac,Kurian:2019nna}. Charm quark transport in the QGP is sensitive to the temperature and transport properties of the medium. The Brownian transport of heavy flavor particles in the QGP along with key experimental observables such as nuclear suppression factor $R_{AA}$ and elliptic flow coefficient $v_2$ has been extensively studied at RHIC and LHC energies~\cite{Cao:2013ita,Cao:2018ews,He:2011qa, Alberico:2013bza,Song:2015sfa,Li:2018izm,Uphoff:2011ad,Das:2013kea,Li:2019wri,Singh:2023smw}. Despite the success of theoretical frameworks in describing the charm quark dynamics in an expanding QGP and explaining the measured data, significant challenges remain. As heavy quarks are predominantly created in the very early stage of collisions, they have the potential to retain imprints of the initially produced matter. However, the effects of the initial stage and the pre-equilibrium evolution of heavy quarks in the fluctuating initial background are often neglected owing to the short duration of this phase. Recent advancements have been made in understanding the heavy quark transport coefficients during the pre-hydrodynamic stage~\cite{Boguslavski:2023fdm, Du:2023izb, Pandey:2023dzz, Liu:2019lac, Carrington:2022bnv, Backfried:2024rub}. These studies suggest that even though the pre-equilibrium phase is short lived, the interaction strength in this stage is large enough to affect heavy flavor observables. Consequently, the potential sensitivity of hard probes to the pre-equilibrium stage in phenomenological studies is gaining attention~\cite{Avramescu:2023qvv,Sun:2019fud,Pooja:2022ojj,Pooja:2024rnn,Avramescu:2024xts}.

In this study, we utilize the impact parameter dependent glasma (IP-Glasma) model~\cite{Schenke:2012wb,Schenke:2012hg} coupled to the viscous hydrodynamic simulation MUSIC (which is short for `MUSCL for Ion Collisions', where MUSCL stands for Monotonic Upstream-centered Scheme for Conservation Laws)~\cite{Schenke:2010nt,Schenke:2010rr,Ryu:2015vwa} and the hadronic transport model Ultrarelativistic Quantum Molecular Dynamics (UrQMD)~\cite{Bass:1998ca}, to model the bulk medium evolution. This framework has been successful in describing a wide range of soft hadronic observables in heavy-ion collisions \cite{Schenke:2020mbo}. The model is based on describing the nuclei, prior to their collision, as two sheets of color glass condensate, which captures the essence of the low momentum-fraction gluons present in the nuclei accelerated to high energies. The gluon field generated by the collision of the energetic nuclei evolves for a short time according to the classical Yang-Mills equations. The energy-momentum tensor of these gluons then serves as the input for viscous hydrodynamics, which takes over the time evolution of the bulk medium. We simulate the evolution of charm quarks starting from the fluctuating IP-Glasma initial stage, followed by their evolution in the hydrodynamically expanding QGP, using the MARTINI (Modular Algorithm for Relativistic Treatment of Heavy Ion Interactions) event generator~\cite{Schenke:2009gb,Young:2011ug}. By incorporating the charm quark dynamics in the initial stage and in the QGP phase, we provide first phenomenological insights into charm quark evolution within the IP-Glasma initial state and its effects on the final state observables.

The article is organized as follows. Sec.~\ref{section II} describes the multi-stage model for charm quark evolution, where we discuss the bulk medium spacetime evolution (Sec.~\ref{II.A}), heavy quark initial production (Sec.~\ref{II.B}), its evolution in the medium (Sec.~\ref{II.C}), and the hadronization mechanism (Sec.~\ref{II.D}). Results and discussions of the study are presented in Sec.~\ref{III}.  Finally, we summarize our findings in Sec.~\ref{IV} and present an outlook.

\section{Charm quark evolution within a multi-stage model }\label{section II}
We adopt multi-stage models to describe the evolution of both the background QGP medium and charm quarks in heavy-ion collisions.

\subsection{Bulk medium evolution}\label{II.A}

All the model parameters for bulk medium evolution are the same as those utilized in Ref.~\cite{Schenke:2020mbo}. The very first stage is modeled by IP-Glasma, which describes the initial production and evolution of color fields due to the impact of two energetic nuclei. IP-Glasma is based on the Color Glass Condensate (CGC) framework~\cite{Gelis:2010nm}. The CGC action can be expressed as~\cite{Krasnitz:1998ns},
 \begin{align}
     S_{\text{CGC}}=\int{d^4 x \Big(-\frac{1}{4}F^a_{\mu\nu}F^{a\, \mu\nu}+J^{a\, \mu}A^a_\mu\Big)},
 \end{align}
 with $J^{a\, \mu}$ the current term that sources soft gluons field $A^a_\mu$, and $F^a_{\mu\nu}$ the field strength tensor, where $a$ denotes the color index. The evolution of the gluon fields is governed by the Classical Yang-Mills (CYM) equations:
 \begin{align}
     [ D_\nu, F^{ \mu\nu}]^a= J^{a\, \mu},
 \end{align}
 with
  \begin{align}
      &D^a_\nu=\partial_\nu-ig A_\nu t^a,\nonumber\\
      & J^{a\, \mu} = \delta^{\mu\pm}\rho^a_{A(B)}\big(x^{\mp}, {\bf x}_\perp \big),
 \end{align}
 where $t^a$ are the SU(3) generators and $\rho^a$ are the color charges. Here, the color current originates from the nucleus $A(B)$ moving along the positive (negative) light-cone direction $x^+ (x^-)$.
 The Glasma distributions, obtained by solving the CYM equations event-by-event, subsequently serve as input to fluid dynamics. We use 140 IP-Glasma events per 10\% centrality.

 The next phase involves the evolution of the QGP medium. We switch from IP-Glasma to a fluid description using the hydrodynamic solver MUSIC at longitudinal proper time $\tau = 0.4$ fm. The state of a fluid can be described by the energy-momentum tensor of the medium.  For dissipative QGP, the ideal form of the energy-momentum tensor is modified with the inclusion of shear tensor $\pi^{\mu\nu}$ and bulk-viscous pressure $\Pi$ as, 
\begin{equation}
     T^{\mu\nu}=\varepsilon u^\mu u^\nu-\Delta^{\mu\nu}(P+\Pi) +\pi^{\mu\nu},
\end{equation}
where $\varepsilon$ and $P$ denote the local energy density and pressure of the QGP, $u^{\mu}$ is the fluid velocity, $g^{\mu\nu}=\text{diag}(1, -1, -1, -1)$ and $\Delta^{\mu\nu}=g^{\mu\nu}-u^{\mu}u^{\nu}$. Second-order viscous hydrodynamics, specifically the DNMR formulation \cite{Denicol:2012cn,Denicol:2014vaa}, incorporating both shear and bulk viscosity, is utilized to describe the evolution of the QGP medium.
The stress tensor and bulk-viscous pressure follow the relaxation-type equations 
\begin{align}
 {\tau_\pi}\dot{\pi}^{\langle\mu\nu\rangle}+\pi^{\mu\nu} &= 2\eta\sigma^{\mu\nu}-\delta_{\pi\pi}\pi^{\mu\nu}\theta+\phi_{7}\pi_{\beta}^{\langle\mu}\pi^{\nu \rangle\beta} \nonumber\\
 &~~~-\tau_{\pi\pi}\pi_{\beta}^{\langle\mu}\sigma^{\nu \rangle\beta}+\lambda_{\pi\Pi}\Pi\sigma^{\mu\nu},\\
 {\tau_\Pi}\dot{\Pi}+\Pi & = - \zeta\theta-\delta_{\Pi\Pi}\Pi\theta
 +\lambda_{\Pi\pi}\pi^{\mu\nu}\sigma_{\mu\nu},
\end{align}
where $\theta=\partial_{\mu}u^{\mu}$ is the scalar expansion rate,  ${\pi}^{\langle\mu\nu\rangle}=\Delta^{\mu\nu}_{\alpha\beta}\pi^{\alpha\beta}$, and $\sigma^{\mu\nu}=\Delta^{\mu\nu}_{\alpha\beta}\partial^{\alpha}u^{\beta}$ with the symmetric, traceless projection operator defined as $\Delta^{\mu\nu}_{\alpha\beta}\equiv\frac{1}{2}(\Delta^\mu_\alpha\Delta^\nu_\beta +\Delta^\mu_\beta\Delta^\nu_\alpha)-\frac{1}{3}\Delta^{\mu\nu}\Delta_{\alpha\beta}$. This framework is implemented numerically in MUSIC. The first-order viscous coefficients, the shear ($\eta$) and bulk ($\zeta$) viscosities, are tuned to data in~\cite{Schenke:2020mbo}. The second-order transport coefficients $\delta_{\pi\pi}, \phi_{7}, \tau_{\pi\pi}, \lambda_{\pi\Pi}, \tau_{\pi}, \delta_{\Pi\Pi}, \lambda_{\Pi\pi}, \tau_{\Pi}$ are expressed in terms of shear and bulk viscosities~\cite{Denicol:2014vaa}. 

The Cooper-Frye procedure~\cite{Cooper:1974mv} is used to determine a freeze out surface from which hadrons are sampled using the iSS sampler~\cite{Shen:2014vra}. Sampled hadrons undergo hadronic scatterings and resonance decays within UrQMD~\cite{Bass:1998ca}. Final hadronic lists from UrQMD are then used to calculate charged hadron observables.

\subsection{Heavy quark production}\label{II.B}

Due to their high mass, heavy quarks are predominantly created in the initial hard collisions. 
The initial spatial distribution of heavy quark production points is obtained from information provided by the IP-Glasma initial state. For each IP-Glasma event, we store the location of the original binary collisions. Then charm/anti-charm quark pairs are sampled using PYTHIA \cite{Sjostrand:2014zea} at the binary collision positions with a probability that depends on the cross-section for charm quark production. 

To account for nuclear shadowing effects, EPS09 nuclear parton distribution functions~\cite{Eskola:2009uj} are utilized. Additionally, isospin effects are considered by sampling $n+n$, $p+p$, and $p+n$ collisions. Only initial charm production is considered in the present analysis, since thermal production of heavy quarks in the medium is negligible, because of the high charm quark mass \cite{Dong:2019unq}.

As the charm spatial and momentum profile is stochastically sampled, we use the same background IP-Glasma event for multiple charm events. For each IP-Glasma event, we sample twenty-five thousand charm events in the 0-10\% centrality bin and fifty thousand charm events in the 30-50\% centrality bin.
    
\begin{figure}
    \includegraphics[width=0.45\textwidth]{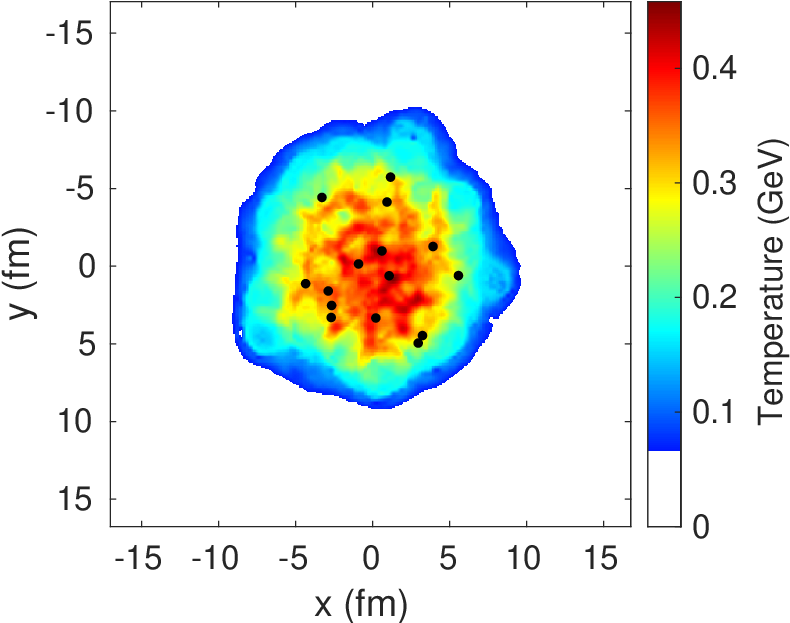}
    \caption{Charm quark locations within the surrounding medium at $\tau = 1.5$ fm at mid-rapidity. }
    \label{fig1}
\end{figure}

We implement a formation time of $\tau_F = 1/m_c$ in the charm quark's rest frame. Here $m_c$ is the mass of a charm quark. Charm quarks are allowed to interact with the medium only after their formation time has passed. Before the formation time is reached, charm quarks/anti-quarks are bound to their original partner via a Cornell potential \cite{PhysRevD.17.3090}. 
We will study the effect of varying the formation time below.  Charm quark positions in the background medium at midrapidity for a single event are shown in Fig.\,\ref{fig1}.

\subsection{Heavy quark evolution in Glasma and QGP}\label{II.C}

The precise estimation of charm quark interaction with the gluon fields in the (3+1)-dimensional space-time geometry in the earliest phase of heavy-ion collisions is a complex task and has not yet been fully developed. Nevertheless, the interesting question is whether the heavy-quark evolution in the initial state has any phenomenological implications, given the short lifetime of this stage. In the present study, we focus on this aspect. As a first step in studying the impact of evolution during the fluctuating initial stage on charm observables, we approximate the initial stage as an evolving QCD medium solely composed of gluons. The evolution of the energy density $\varepsilon$ in the initial stages is described by the IP-Glasma model. We assume that the system is locally thermalized to extract an effective 
temperature, from which we can compute the charm quark drag and diffusion coefficients.

These then enter the calculation of the in-medium dynamics of the heavy quarks. In both the Glasma and QGP, it is modeled as Brownian motion within MARTINI, where its interaction strength with the medium is quantified via the drag and diffusion coefficients. The momentum change $dp_i$ of the charm quark over a time interval $dt$ can be described using the Langevin equations in the local rest frame of the medium as~\cite{Moore:2004tg,Das:2013kea},
\begin{align}
   &dp_i=-\eta(|{\bf p}|) p_i\, dt+ \xi_i ({\bf p})\, {dt},\nonumber\\
   &\langle \xi_i (t)\xi_j (0)\rangle = \Big(\delta_{ij}-\frac{p_ip_j}{|{\bf p}|^2}\Big)\,\kappa_T(|{\bf p}|)+\frac{p_ip_j}{|{\bf p}|^2}\,\kappa_L (|{\bf p}|),
\end{align}
with $\eta(p_i)$ the drag coefficient, $\kappa_T$ the transverse and $\kappa_L$ the longitudinal momentum diffusion coefficients. The term $-\eta(|{\bf p}|) p_i$ is the drag force that quantifies the average momentum transfer of the charm quark due to the interaction, and $\xi_i$ represents the stochastic force acting on the charm quark.

To study the evolution of a charm quark in the medium, first we obtain the energy-density and fluid four-velocity at the space-time location of the charm quark from IP-Glasma or MUSIC. In the IP-Glasma phase, temperature is obtained from this energy-density assuming the ideal gas equation of state with gluon degrees of freedom. In the QGP phase, in MUSIC, we use the lattice QCD equation of state matched with a hadronic resonance gas at low temperatures \cite{HotQCD:2014kol}. The next step is to boost the heavy quark momentum to the local rest frame of the medium and to estimate the change in momentum $dp_i$ during the time step $dt$. After modifying the momentum, we boost back the charm quark momentum to the lab frame and update its position after $dt$.

The detailed transport of heavy quarks in the QCD medium depends on their transport coefficients in the respective medium. One of the latest advances in this direction is the first $N_f=2+1$ lattice estimation of the spatial diffusion coefficient $D_s$ for the temperature range $195$ MeV $< T < 352 $ MeV with dynamic light quarks corresponding to a pion mass of $320$ MeV~\cite{Altenkort:2023oms}. In this work, we parametrize the spatial diffusion coefficient and extrapolate it to temperatures outside this range (with fluctuating initial conditions some hot spots can reach much higher temperature, especially at the earliest times). A more recent study extends the $D_s$ estimation to higher temperatures~\cite{HotQCD:2025fbd}, but it is consistent with the previous result and our extrapolation. 

\begin{figure*}
    \centering
    \includegraphics[width=0.45\textwidth]{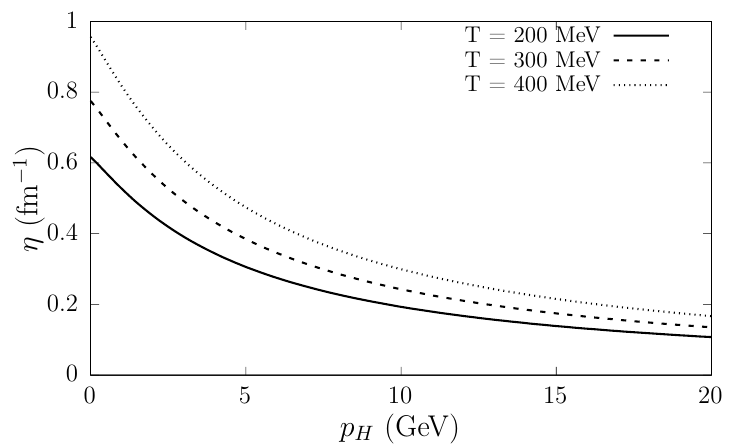}
    \includegraphics[width=0.45\textwidth]{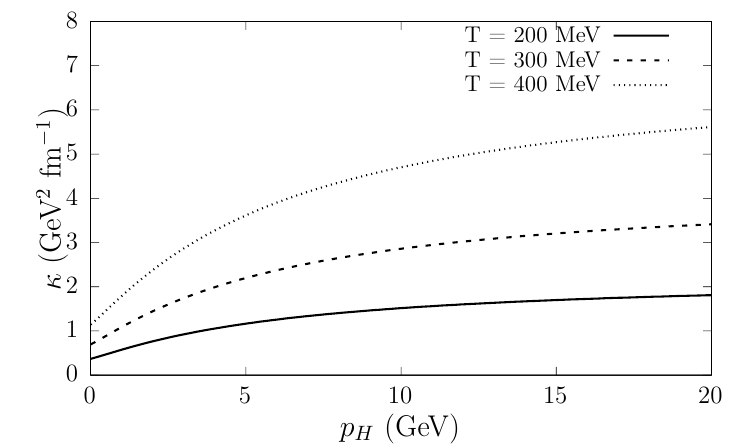}
    \caption{Momentum dependence of drag (left panel)
    and diffusion (right panel)
    coefficients}
    \label{fig2}
\end{figure*}

The drag and diffusion coefficients are obtained from $D_s$ at zero momentum using the Einstein relations $\kappa = 2T^2/D_s$ and $\eta = T/(m_cD_s)$. Further, we parametrize the momentum dependence of $\eta$ and $\kappa$.  The parametrization is inspired by the perturbative QCD calculation of charm quark transport coefficients~\cite{GolamMustafa:1997id,Kurian:2020orp}.  The momentum dependence of $\eta$ and $\kappa_T=\kappa_L=\kappa$ is shown in Fig.~\ref{fig2}.
It is observed that the heavy quark drag coefficient decreases as the heavy quark momentum increases, while the diffusion coefficients show the opposite trend. With increasing heavy quark momentum the charm quark diffusion coefficient rises and eventually saturates at high momentum. We will explore the effect of including the momentum dependence on observables later in this work.

It is important to note that with the correct choice of critical temperature, the spatial diffusion coefficients obtained from $N_f=0$ lattice calculations~\cite{Banerjee:2022uge}, are consistent with the results of the 2 + 1 flavor lattice calculations~\cite {Altenkort:2023eav}. Hence we use the same $D_s$ parameterization in both the Glasma and the QGP phases.

\subsection{Heavy quark hadronization}\label{II.D}
Charm quarks are evolved in the medium to the hadronization temperature $T_{h} = 165$ MeV. At $T = T_{h}$, a modified color evaporation model is first employed to determine whether a charm quark forms a charmonium state with another charm quark. Additionally, recombinant charmonium production is considered for all the remaining charm quark/anti-quark pairs. Details of the algorithm used for charmonium production is given in Ref.~\cite{Young:2011ug}.

We employ a combination of fragmentation and heavy-light coalescence mechanisms for hadronizing charm quarks that remain after charmonium production. Experimental results have indicated the necessity of a coalescence approach in the low $p_T$ regime. In the coalescence
model, the probability that a heavy-flavor quark will coalesce with a neighboring light-flavor quark (or quarks) to form a hadron is described by the momentum-space Wigner function $f^W_H$. The momentum spectrum of the produced hadron (D-mesons and baryons) can be written as~\cite{Zhao:2023nrz}, 
\begin{align}
    \frac{dN_H}{d^3p_H}=&\int  \prod_{i=1}^{N}  {d^3p_i}\frac{dN_i}{d^3p_i}f^W_H({\bf p}_1, {\bf p}_2, ...,  {\bf p}_N) \notag\\ &~~~~~~~~~\times \delta \big({\bf p}_H-\sum_{i=1}^{N}{\bf p}_i\big),
\end{align}
where ${\bf p}_H$ is the momentum of the heavy flavor hadron $H$ and ${\bf p}_i$ are the momenta of the constituent quarks. Here, $dN_i/d^3p_i$ is the momentum distribution of the $i-$th constituent quark in the recombined meson or baryon. The conservation of 3-momentum is ensured by the $\delta-$function.

\begin{figure}
    \centering
    \includegraphics[width=0.45\textwidth]{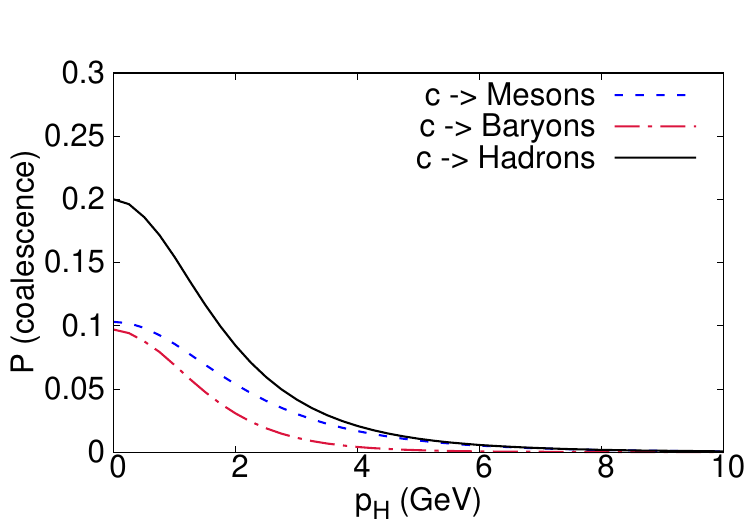}
    \caption{Probability of a charm quark coalescing into a meson, baryon, or hadron.}
    \label{fig:Coalescence_prob}
\end{figure}

In this work, we use the instantaneous coalescence model~\cite{Cao:2015hia}. The momentum space Wigner function for an S-wave meson takes a Gaussian form and can be written as,
\begin{align}
    f^W_{H=M}=N_{coal} g_M \frac{(2\sqrt{\pi }\sigma)^3}{V}\exp \Big(-p_r^2\sigma^2\Big),
\end{align}
where the subscript $M$ denotes mesons, $N_{coal}$ is a normalization factor, and $g_M$ is the degeneracy factor of the mesons. The quark distributions is assumed to be uniform in the volume $V$. The width parameter of the Gaussian distribution is defined as $\sigma=1/\sqrt{\mu\omega}$ where $\mu = m_cm_{u/d}/(m_c+m_{u/d})$ is the reduced mass of the heavy and light quark system. For quantitative estimation, we consider the thermal mass of light quarks as $m_{u/d}= 300$ MeV. The angular frequency of the oscillator $\omega$ is taken to be $0.106$ GeV, as tuned in Refs.~\cite{Cao:2013ita, Oh:2009zj}. The relative momentum in the two-body center-of-mass frame is
\begin{align}\label{rel:mom}
    {\bf p}_r= \frac{1}{E'_1+E'_2}(E'_2{\bf p}'_1-E'_1{\bf p}'_2).
\end{align}
For baryons (B) such as $\Lambda_c, \Sigma_c$, $\Xi_c$ and $\Omega_c$, the Wigner function can be extended to a three-particle system by initially combining two particles and then using the center of mass of this pair to combine with the third particle as~\cite{Cao:2015hia},
\begin{align}
    f^W_{H=B}=N_{coal} g_B \frac{(2\sqrt{\pi})^6 (\sigma_1 \sigma_2)^3}{V^2}\exp{\Big(-p_{r\,1}^2\,\sigma_1^2-p_{r\,2}^2\, \sigma_2^2\Big)},
\end{align}
where $g_B$ is the baryon degeneracy factor and $\sigma_i=1/\sqrt{\mu_i\omega}$. The parameters $\mu_1$ and $\mu_2$ describe the reduced mass of a two-particle and three-particle system, respectively. The relative momentum ${\bf p}_{r\,1}$ of two particles in the center-of-mass frame takes the same form as that of Eq.~(\ref{rel:mom}) and relative momentum ${\bf p}_{r\,2}$ with the third particle can be defined as,
\begin{align}
    {\bf p}_{r\,2}= \frac{1}{E'_1+E'_2+E'_3}\Big(E'_3({\bf p}'_1+{\bf p}'_2)-(E'_1+E'_2){\bf p}'_3\Big).
\end{align}
We calculate the total probability for a charm quark of momentum $p_H$ to hadronize by coalescence and the probability that the coalesced hadron is a meson. There is an overall normalization factor $N_{coal}$ that determines the probability that charm quarks at rest in the local rest frame of the medium hadronize by coalescence. We fix $N_{coal} = 0.2$ to get a good agreement with the experimental data. The coalescence probabilities are shown in Fig.\,\ref{fig:Coalescence_prob}. 
For every charm quark, a random number between 0 and 1 is generated. If this value is less than the probability of coalescence, the heavy quark undergoes hadronization by coalescence. If it is greater than the probability, then it hadronizes by fragmentation. 

For fragmentation, charm quarks are rotated to the $z$-axis and their four momenta are written in lightcone coordinates. A charm quark with momentum $p_+$ in lightcone coordinates in the rotated frame fragments into a D-meson with momentum $zp_+$. A random transverse kick is sampled from a Gaussian of width 0.335 GeV. The kick is applied to the D-meson before rotating the four-momentum back to the original frame. The fragmentation factor $z$ is sampled from the Peterson fragmentation function~\cite{Peterson:1982ak},
\begin{align}
    f(z)\propto \frac{1}{z\Big(1-\frac{1}{z}-\frac{\bar{\varepsilon}}{(1-z)}\Big)^2},
\end{align}
with the parameter $\bar{\varepsilon}=0.15$. We put a lower cutoff on $z$ such that $z > m_D/p_+$.
 
\section{Results and Discussions}\label{III}

We analyze the nuclear suppression factor $R_{AA}$ and the elliptic flow coefficient $v_2$ of D mesons in Pb+Pb collision at $\sqrt{s_{NN}}= 5.02$ TeV within the multi-stage model. The energy loss of charm quarks in heavy-ion collision can be quantified in terms of the nuclear suppression factor as,
\begin{align}
    R_{AA} = \frac{1}{N_{\rm coll}}\frac{dN_{AA}/dp_T}{dN_{pp}/dp_T}, 
\end{align}
where $dN_{AA}/dp_T$ is the D meson yield in the nucleus-nucleus collision for a given centrality range and $dN_{pp}/dp_T$ denotes the D meson yield in minimum bias proton-proton collisions. Here, $N_{\rm coll}$ is the average number of binary collisions in the considered centrality class of the nucleus-nucleus collision.

The elliptic flow coefficient $v_2$ is written in terms of complex $Q_2$ vectors, which are defined as
\begin{equation}
    Q_2 = \frac{1}{N}\sum_{k=1}^{N} e^{2i\phi_k},
\end{equation}
where $\phi_k$ is the azimuthal angle of the $k-$th particle. Here $N$ is the total number of particles of interest. The ALICE collaboration used the scalar product (SP) definition of D-meson $v_2$ \cite{ALICE:2020iug, Voloshin:2008dg, Luzum:2012da}
\begin{equation}\label{eq:v2SP}
    v_2\{{\rm SP}\} =  \left.\langle Q_{2D} Q^*_{2A}\rangle\middle/\sqrt{\frac{\langle Q_{2A}Q^*_{2B}\rangle\langle Q_{2A}Q^*_{2C}\rangle}{\langle Q_{2B}Q^*_{2C}\rangle}}\right.,
\end{equation}
where $Q_{2D}$ is the $Q_2$ vector of D mesons and $Q_{2A}$, $Q_{2B}$ and $Q_{2C}$ are $Q_2$ vectors of subevents $A$, $B$ and $C$. The ALICE collaboration uses scintillator detectors (V0) in the backward and forward direction for subevents $A$ and $B$ and the Time Projection Chamber (TPC) at midrapidity for subevent C. The V0 detectors measure energy deposition while the TPC can measure charged particle yields. In our model, the bulk hydrodynamic background is boost invariant, and we use mid-rapidity charged hadron $Q_2$ vectors $Q_{2h}$ as a substitute for all the subevents $A, B$, and $C$. Then Eq.\,\eqref{eq:v2SP} simplifies to
\begin{equation}\label{eq:v2_def}
    v_2\{{\rm SP}\} = \frac{\langle Q_{2D}Q^*_{2h}\rangle}{\sqrt{\langle Q_{2h}Q^*_{2h}\rangle}},
\end{equation}
which we use to calculate D-meson $v_2$.

Two particle azimuthal correlations have also been suggested to be sensitive to pre-equilibrium evolution \cite{Avramescu:2024xts}. The calculation of this observable within our model will be given in \cite{plannedpaper}.

\begin{figure*}
    \centering
    \includegraphics[width=0.45\textwidth]{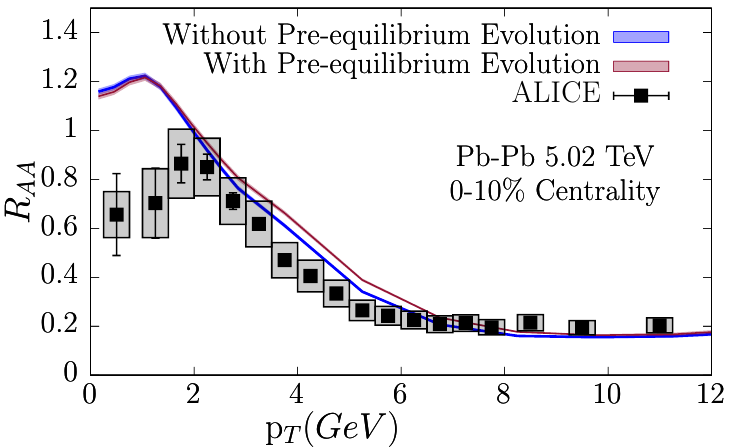}
    \includegraphics[width=0.45\textwidth]{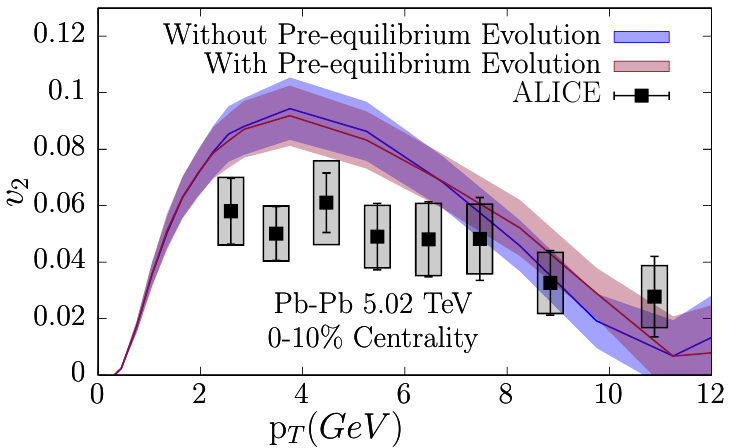}
    \includegraphics[width=0.45\textwidth]{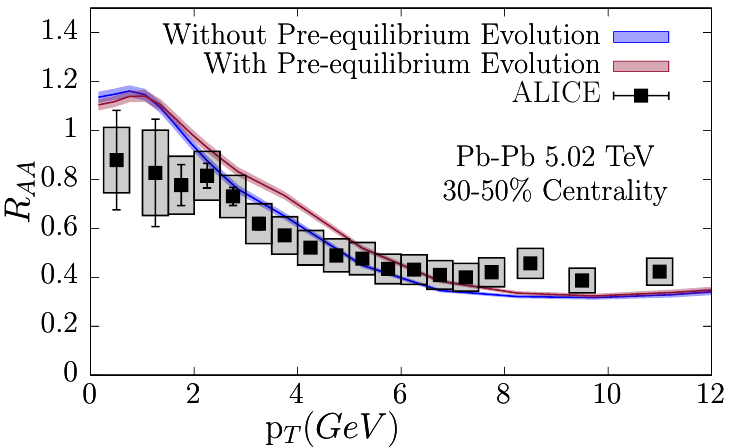}
     \includegraphics[width=0.45\textwidth]{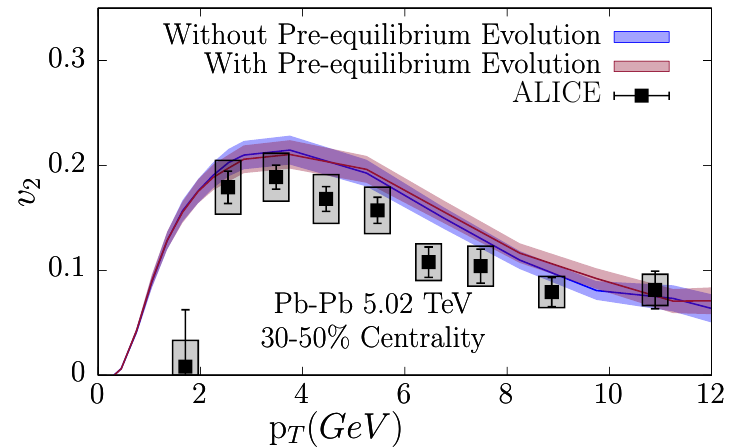}
    \caption{D-meson $R_{AA}$ and $v_2$ with and without charm quark pre-equilibrium evolution at different centralities. Experimental data taken from~\cite{ALICE:2021rxa,ALICE:2020iug}.}
    \label{preequilibrium}
\end{figure*}

\begin{figure}
    \centering
    \includegraphics[width=0.45\textwidth]{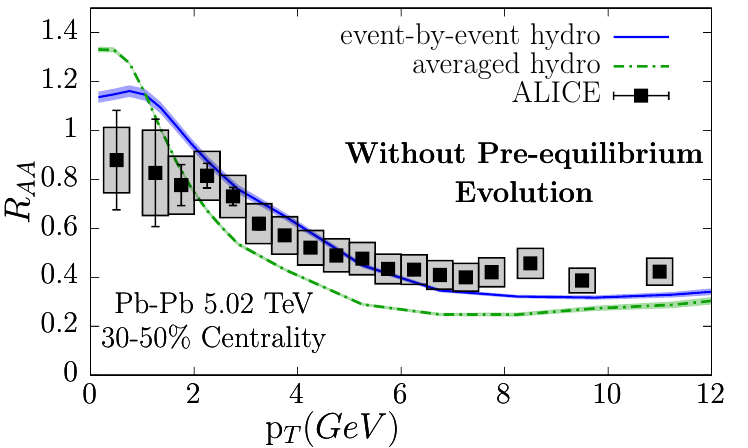}
    \includegraphics[width=0.45\textwidth]{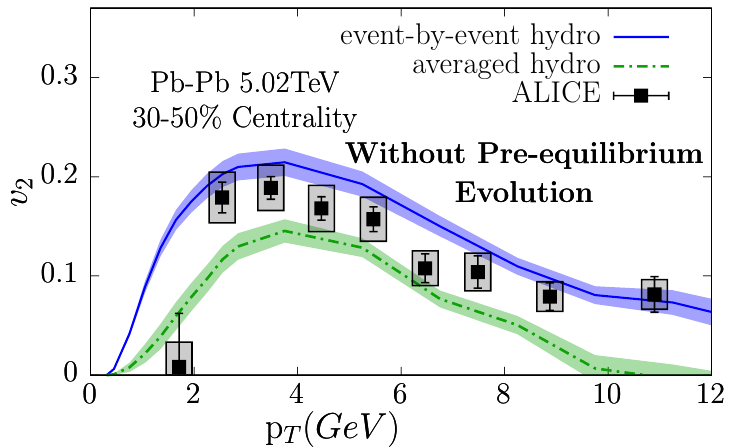}
    \caption{ Impact of event-by-event (ebe) versus averaged hydro background on charm observables. Experimental data taken from~\cite{ALICE:2021rxa,ALICE:2020iug}. }
    \label{figcomparison}
\end{figure}

\subsection{Impact of charm quark pre-equilibrium evolution on observables}\label{III.B}
As charm quarks propagate through the Glasma, they undergo energy loss and momentum broadening. Despite the shorter lifetime of the pre-equilibrium phase, the Glasma contributes to the heavy quark transport and thereby the momentum broadening. Recent studies~\cite{Carrington:2021dvw, Carrington:2022bnv} have found that the pre-equilibrium and equilibrium contributions to the total transverse momentum broadening of a charm quark for a one-dimensional boost-invariant hydrodynamical expansion are comparable with $(\Delta p_T^2\vert^{\text{non-eq}})/(\Delta p_T^2\vert^{\text{eq}})=0.93$. In our analysis using the IP-Glasma+MUSIC+UrQMD framework, we obtain a result consistent with this, namely
\begin{align}
\frac{\Delta p_T^2\vert^{\text{non-eq}}}{\Delta p_T^2\vert^{\text{eq}}}\approx 0.8-1,
\end{align}
depending on the centrality.

To evaluate the phenomenological significance of this pre-equilibrium contribution, we calculate the D-meson observables by including charm quark evolution during the pre-equilibrium phase. Fig.~\ref{preequilibrium} shows the resulting $R_{AA}$ and $v_2$ of D-mesons, comparing scenarios with and without pre-equilibrium charm quark energy loss for different centralities. We consider the formation time of the charm quark as $\tau_F=1/m_c$ (in the quark's rest frame), where $m_c$ is the mass of the charm quark. 
In the scenario that includes pre-equilibrium evolution (represented by the red curve), the charm quark undergoes energy loss and momentum broadening during the Glasma phase until $\tau=0.4$ fm, after which they propagate through the QGP. Conversely, in the absence of pre-equilibrium effects (denoted by the blue curves), the charm quark evolution is described by the equation of motion under the zero-temperature Cornell potential until their formation, or until $\tau = 0.4$ fm, whichever is later. After this time, similar to the first case, the charm quark is transported through the QGP medium. The charm quark pre-equilibrium evolution slightly modifies the final D-meson $R_{AA}$ in the momentum range $2$ GeV $<p_T< 8$ GeV for both centralities. The elliptic flow is little affected by the charm quark pre-equilibrium evolution. It is interesting to note that the considered final charm observables are not sensitive to the pre-equilibrium evolution of the charm quark, despite the significant contribution to the momentum broadening. 

\begin{figure*}
    \centering
    \includegraphics[width=0.45\textwidth]{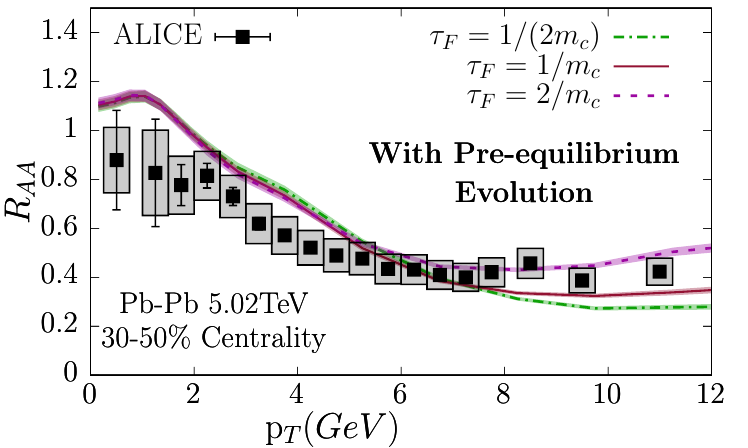}
    \includegraphics[width=0.45\textwidth]{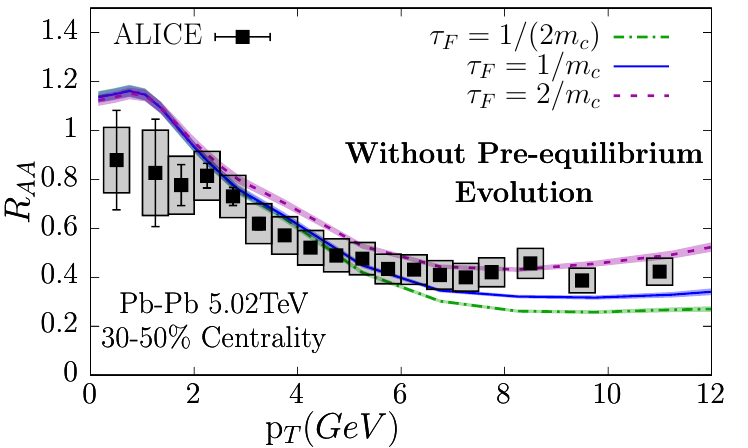}
    \includegraphics[width=0.45\textwidth]{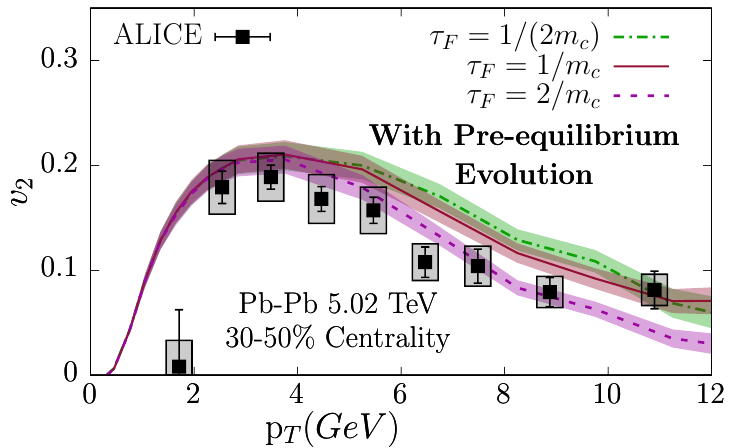}
    \includegraphics[width=0.45\textwidth]{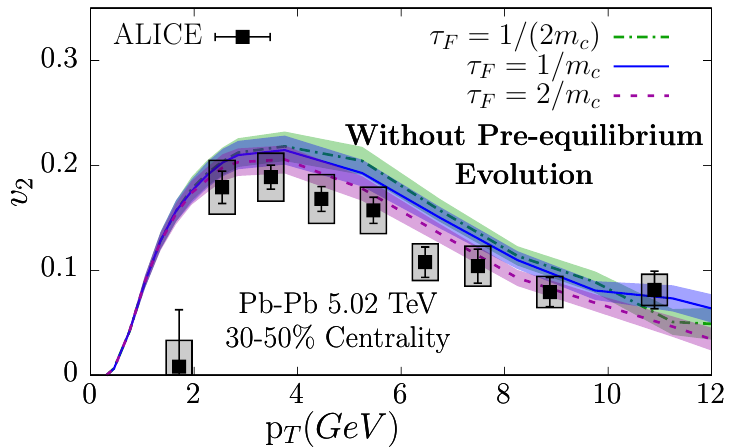}
    \caption{Sensitivity of D-meson $R_{AA}$ and $v_2$ on the formation time of charm quarks in the $30-50\%$ centrality bin. Experimental data taken from~\cite{ALICE:2021rxa,ALICE:2020iug}.}
    \label{formation time}
\end{figure*}

The enhancement of $R_{AA}$ with an added Glasma interaction phase may seem counter-intuitive. Quarks indeed face more drag as a result of this added interaction, which reduces their momentum. But this is more than compensated by the high-momentum kicks received at large temperatures in the Glasma phase. These kicks are encoded in the diffusion term. They are primarily responsible for large momentum broadening and they also enhance the momentum of low-momentum charm quarks.

Charm quarks are isotropically oriented after their production from initial hard scatterings. Large diffusion in the Glasma phase effectively scrambles these charm quarks but the final result is that they are still isotropically oriented. At this stage, background medium flow has not developed, and charm quark momenta continue to stay uncorrelated with the background. So while their momentum broadening may be large, there is no net phenomenological effect on $v_2$.

The small effect in $R_{AA}$ may not be discernible in experiments, as the size of the effect is smaller than systematic uncertainties in the model. We quantify some of these uncertainties in the remainder of this section.

\subsection{Fluctuating vs smooth initial conditions}\label{III.A}
In this subsection, we emphasize the importance of the event-by-event simulation employed in our study. The $R_{AA}$ and $v_2$ of D-mesons for the centrality $30-50\%$ from the IP-Glasma+MUSIC+UrQMD framework without any Glasma energy loss are depicted in Fig.~\ref {figcomparison}. These results are compared with those obtained using a smooth hydrodynamic background to highlight the effects of a fluctuating background. The smooth profile is generated by using an initial state obtained by averaging all IP-Glasma initial states in the 30-50\% centrality bin. The IP-Glasma initial states are rotated by their spatial eccentricity angle before averaging. All the other parameters are kept the same. We see that both the $R_{AA}$ and the $v_2$ are enhanced for the event-by-event fluctuating hydro background. This highlights the importance of using a realistic background when calculating charm observables.
\subsection{Uncertainty from formation time and coalescence probability}\label{III.C}
\begin{figure}
    \centering
    \includegraphics[width=0.45\textwidth]{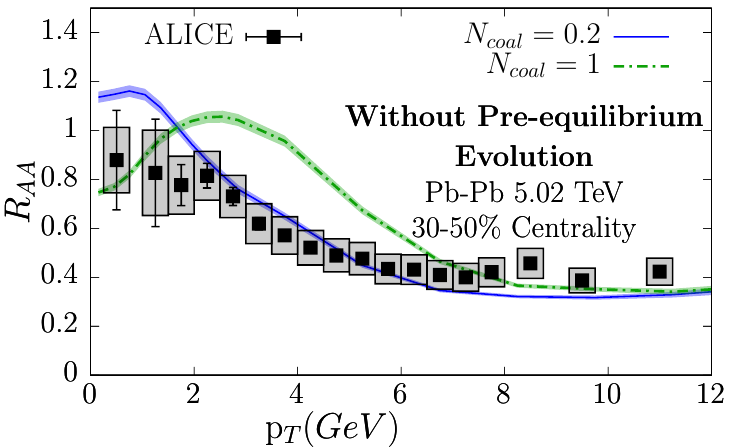}
    \includegraphics[width=0.45\textwidth]{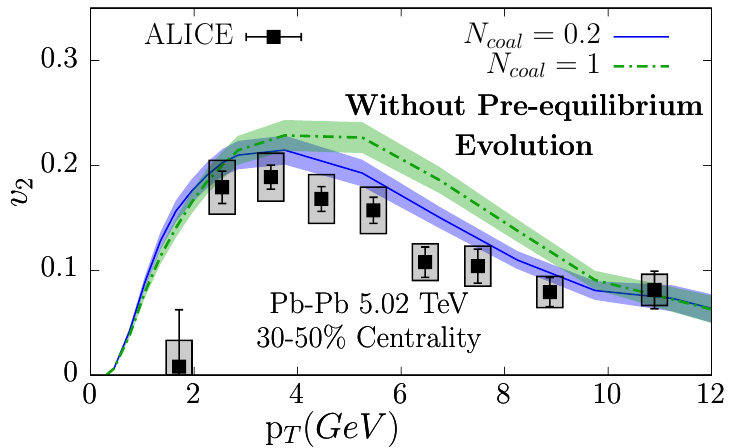}
    \caption{D-meson $R_{AA}$ (top) and $v_2$ (bottom) for different values of coalescence normalization parameter $N_{coal}$. Experimental data taken from~\cite{ALICE:2021rxa,ALICE:2020iug}.} 
    \label{fig:coalN}
\end{figure}
We chose a formation time $\tau_F = 1/m_c$ in the quark's rest frame. In the lab frame, in Minkowski coordinates, this translates to $t_F = E/m_c^2$. This is an important parameter as it decides how much interaction the charm quark undergoes in the earliest stage of the Glasma. For more energetic charm quarks, formation times may be larger than the hydro start-time. In that case, formation time also decides how much of the hot QGP is seen by charm quarks. We do simulations after halving and doubling the formation times. The sensitivity of D-meson $R_{AA}$ and $v_2$ at 30-50\% centrality on the formation time of the charm quark is shown in Fig.\,\ref{formation time}.  The results indicate that formation time has a notable effect on both $R_{AA}$ and $v_2$ in the high $p_T$ region. The observation holds true with and without pre-equilibrium evolution of the charm quarks. For a shorter formation time, $R_{AA}$ is suppressed more in the high $p_T$ regime, whereas $v_2$ is enhanced. This is because charm quarks see more of the medium and lose more energy. We observe a similar effect in the 0-10\% centrality class.

The normalization factor $N_{coal}$ quantifies the coalescence probability of charm quarks in the medium. Fig.~\ref{fig:coalN} compares results for $N_{coal} = 0.2$ and $N_{coal} = 1$, with $N_{coal} = 1$ corresponding to full coalescence in the static limit (i.e., at zero momentum). Reducing $N_{coal}$ from 1 to 0.2 primarily modifies the low-$p_T$ region of $R_{AA}$, with a smaller but noticeable impact on $v_2$. We note that the hadronization model employed in the present analysis is relatively simple. 
Nevertheless, we do not expect a more sophisticated hadronization model to significantly alter how pre-equilibrium dynamics affect the studied observables, leaving our conclusions unaffected. A comprehensive comparison of various hadronization models for D-meson and quarkonia is provided in Ref.~\cite{Zhao:2023nrz}. The analysis of the impact of various hadronization models on the observables is left for a future study.

\subsection{Uncertainty from charm quark transport coefficients}\label{III.D}
 \begin{figure}
    \centering
    \includegraphics[width=0.45\textwidth]{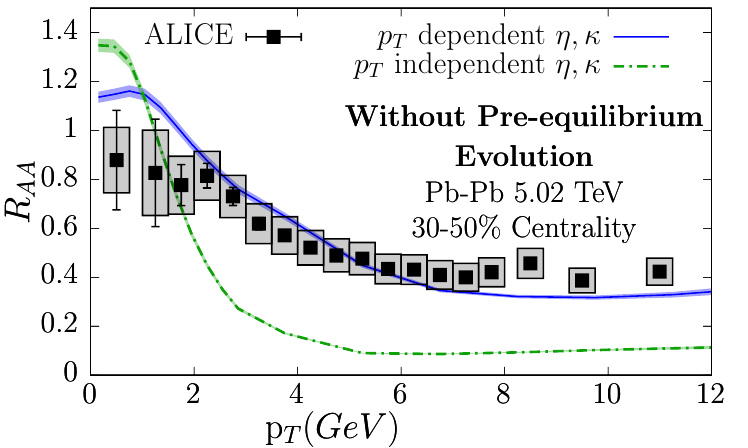}
    \includegraphics[width=0.45\textwidth]{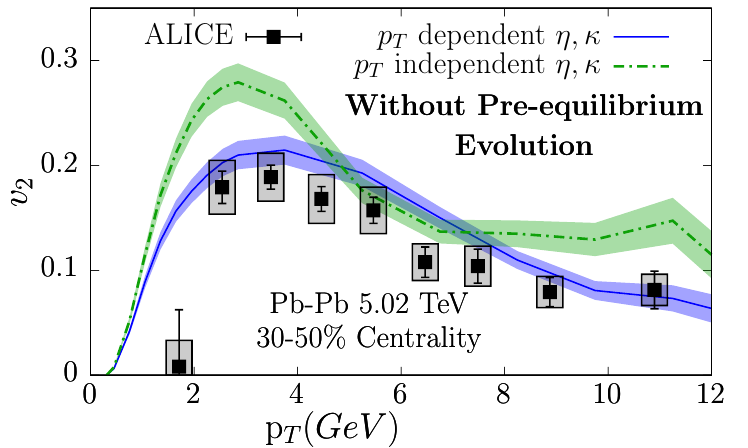}
    \caption{Effect of $p_T$ dependent charm quark transport coefficients on D-meson observables. Experimental data taken from~\cite{ALICE:2021rxa,ALICE:2020iug}.}
    \label{figpt}
\end{figure}
\begin{figure*}
    \centering
    \includegraphics[width=0.45\textwidth]{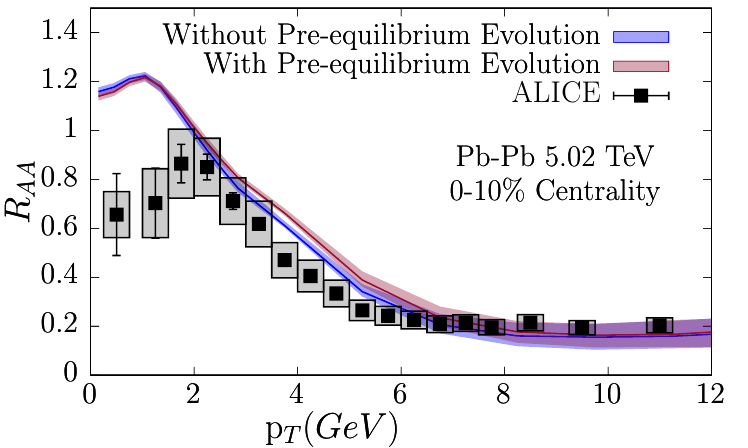}
    \includegraphics[width=0.45\textwidth]{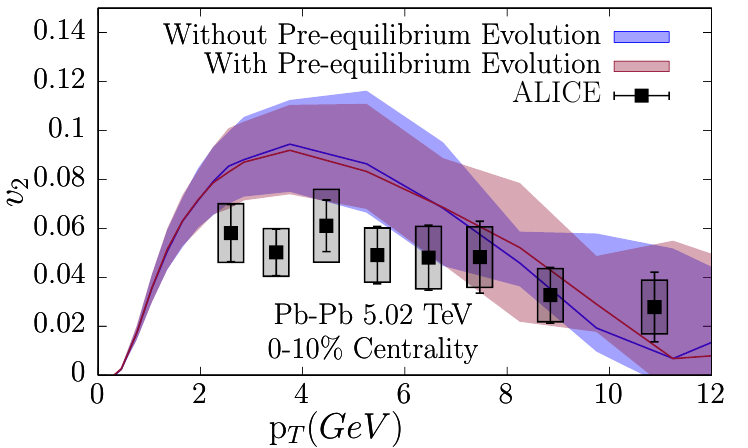}
    \includegraphics[width=0.45\textwidth]{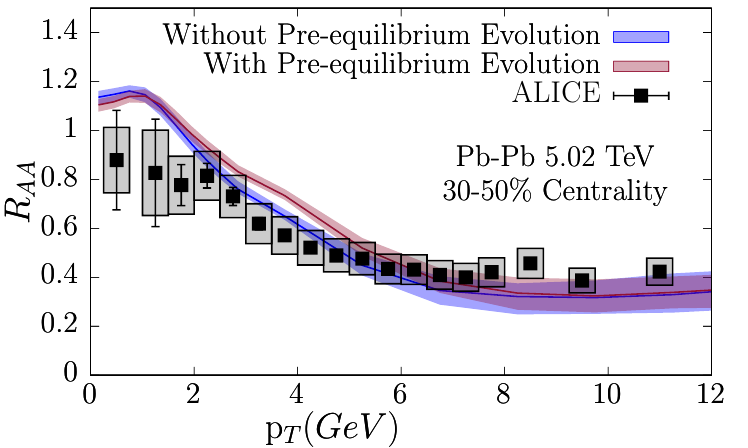}
    \includegraphics[width=0.45\textwidth]{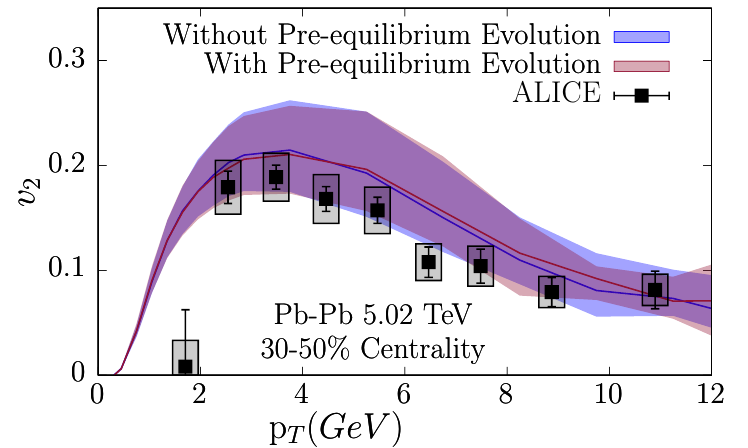}
    \caption{Impact of the lattice QCD uncertainty of the temperature dependence of spatial diffusion coefficient on observables. Experimental data taken from~\cite{ALICE:2021rxa,ALICE:2020iug}.}
    \label{figuncertanity}
\end{figure*}

In general, the heavy quark transport coefficients depend on its momentum and the temperature of the surrounding medium. Heavy quark transport coefficients can be estimated within leading order perturbative QCD while including the elastic and inelastic scattering processes of heavy quarks in the medium. However, attempts to use the Debye mass as an infrared regulator in the gluon propagator for t-channel interactions, along with a fixed coupling constant in perturbative QCD matrix elements, have shown limitations in accurately describing experimental observables related to heavy flavor particles. Several efforts have been made in the estimation of the heavy quark diffusion coefficient using various non-perturbative models and parameters. We refer to Ref.~\cite{Cao:2018ews} and references therein for details. Most of the approaches struggle to simultaneously describe the key observables associated with heavy quarks, the heavy meson nuclear modification factor $R_{AA}$, and elliptic flow $v_2$ (commonly referred to as “heavy quark $R_{AA}$ and $v_2$ puzzle”). Many studies have introduced a K-factor to scale the scattering cross section of heavy quark interactions in the medium, which can be adjusted until the model describes the experimental data. It is important to emphasize that the current analysis utilizes the recent $N_f=2+1$ flavor lattice QCD result for the temperature dependence of the charm quark spatial diffusion coefficient. The necessity of including the momentum dependence of the charm quark transport coefficients in order to accurately describe the $R_{AA}$ and $v_2$ of D mesons is illustrated in Fig.~\ref{figpt}. The effect of the $p_T$ dependence of the energy loss of the charm quark in the medium is clear, especially at the higher momentum regime of $R_{AA}$. 

Though we can get a good agreement with data by utilizing the temperature and momentum dependence of transport coefficients, discrepancies remain in certain momentum ranges. These deviations primarily arise from uncertainties in the precise temperature and momentum dependence of heavy quark transport coefficients. To account for the former, we consider the uncertainty band on the temperature-dependent $N_f=2+1$ flavors lattice result~\cite{Altenkort:2023oms}. The consideration of the lattice uncertainty is crucial, particularly because we extrapolate the lattice data for high temperatures. The inclusion of the $2\pi TD_s$ uncertainty band modifies the uncertainty of both $R_{AA}$ and $v_2$ as shown in Fig.~\ref{figuncertanity}. The lattice QCD uncertainty has a visible impact on the errors of the observables, the effect being more prominent for $v_2$, especially for central collisions. For $R_{AA}$, we still observe a small difference between the cases with and without charm quark pre-equilibrium evolution in the intermediate momentum regimes. However, the effect of pre-equilibrium evolution is insignificant at high $p_T$ as the $R_{AA}$ bands overlap each other. Notably, most of the data points are within the uncertainty band when including the uncertainty from the lattice calculation. 

Lattice QCD calculations are performed in the static limit, where the momentum of the heavy quark is effectively zero, making it challenging to quantify the uncertainties associated with the momentum dependence of the transport coefficients.  While parametrized momentum-dependent forms of these coefficients can be derived from various theoretical models, a more precise approach to determine the momentum dependence involves conducting a systematic model-to-data Bayesian analysis, as discussed in Ref.~\cite{Xu:2017obm}. However, such an analysis lies beyond the scope of the present study.

\section{Summary and Outlook}\label{IV}
We studied the evolution of charm quarks from the very initial state of heavy-ion collision within a multi-stage hybrid framework. The IP-Glasma model was used to describe the fluctuating initial state, followed by MUSIC for the hydrodynamic evolution of the QGP, and UrQMD for final hadronic interactions. One of the key questions explored in this work is whether the charm quark can serve as an efficient probe of the initial stage. The exact solution to this aspect requires a detailed understanding of how charm quarks interact with the non-Abelian gauge fields, which is a highly complex and challenging task. In the present study, we provide a qualitative estimation of the effect of the evolution of charm quarks during the pre-equilibrium phase on charm meson observables, approximating this early stage as a thermalized system composed predominantly of gluons. The QGP evolution is modeled hydrodynamically using MUSIC, and the interaction between charm quarks and the bulk medium is quantified using recent lattice QCD estimates. We found that while the pre-equilibrium phase makes a significant contribution to the momentum broadening of the charm quark, $({\Delta p_T^2\vert^{\text{non-eq}})/(\Delta p_T^2\vert^{\text{eq}}})\approx 0.8-1$, its effect on the considered final state observables is small. The pre-equilibrium evolution slightly enhances the D-meson $R_{AA}$ in the momentum range $2$ GeV $<p_T< 8$ GeV, while its effect on $v_2$ is minimal. Furthermore, the size of the effect is comparable to the size of the systematic uncertainties inherent to the model.

A precise determination of the interaction strength between heavy quarks and Glasma fields is essential for moving beyond the approximation of the Glasma as a locally thermalized gluon bath. A recent study~\cite{Pandey:2023dzz} has extracted the diffusion coefficient of heavy quarks in the presence of highly occupied, non-Abelian gauge fields.  We aim to explore this aspect further in a forthcoming study. In addition, a radiation-improved Langevin model can be used to calculate the observables~\cite{Cao:2013ita} within the developed framework. This will be particularly relevant in view of the recent observation of the dead-cone effect~\cite{ALICE:2021aqk} in QCD.


\section*{Data Availability}
The data that support the findings of this article are
openly available \cite{data_2509_18647}.

\section*{Acknowledgments}
This research used the resources of the National Energy Research Scientific Computing Center, a DOE Office of Science User Facility
supported by the Office of Science of the U.S. Department of Energy
under Contract No. DE-AC02-05CH11231 using NERSC awards NP-ERCAP0033578 and NP-ERCAP0032155. This work is supported by the U.S. Department of Energy, Office of Science, Office of Nuclear Physics, under DOE Contract No.~DE-SC0024347 (M.S.) and~DE-SC0012704 (B.P.S.) and within the framework of the Saturated Glue (SURGE) Topical Theory Collaboration (B.P.S.). M.S. is also supported by the Vanderbilt University. M.K. acknowledges the Special Postdoctoral Researchers Program of RIKEN, the Faculty Research Scheme (FRS project number: MISC 0240) at IIT (ISM) Dhanbad, and the Department of Science and Technology (DST), Govt. of India, for the INSPIRE-Faculty award (DST/INSPIRE/04/2024/001794). S.J. and C.G. are supported by the Natural Sciences and Engineering Research Council of Canada under grant numbers SAPIN-2024-00026 and SAPIN-2020-00048, respectively.


\bibliography{ref}{}

 \appendix

\end{document}